\documentclass[pra,aps]{revtex4}
\usepackage{graphicx}
\newcommand{\be}{\begin{equation}}
\newcommand{\ee}{\end{equation}}
\newcommand{\ba}{\begin{eqnarray}}
\newcommand{\ea}{\end{eqnarray}}
\newcommand{\ban}{\begin{eqnarray*}}
\newcommand{\ean}{\end{eqnarray*}}

\newcommand{\braket}[2]{\mbox{$ \langle #1 | #2 \rangle $}}
\newcommand{\sandwich}[3]{\mbox{$ \langle #1 | #2 | #3 \rangle $}}
\newcommand{\ket}[1]{\mbox{$ | #1 \rangle $}}

\newcommand{\demi}{\frac{1}{2}}

\newcommand{\one}{\leavevmode\hbox{\small1\normalsize\kern-.33em1}}

\begin{document}
\date{\today}
\title{Two independent photon pairs versus four-photon entangled states in parametric down conversion}

\author{Hugues de Riedmatten, Valerio Scarani, Ivan Marcikic, Antonio Ac\'{\i}n$^\dagger$,
Wolfgang Tittel, Hugo Zbinden and Nicolas Gisin}

\affiliation{Group of Applied Physics, University of Geneva, 20,
rue de l'Ecole-de-M\'edecine, CH-1211 Geneva 4, Switzerland\\
$^\dagger$ Presently at: Institut de Ci\`encies Fot\`oniques,
Jordi Girona 29, 08034 Barcelona, Spain }

\begin{abstract}
We study the physics of four-photon states generated in
spontaneous parametric down-conversion with a pulsed pump field.
In the limit where the coherence time of the photons $t_c^{ph}$ is
much shorter than the duration of the pump pulse $\Delta t$, the
four photons can be described as two independent pairs. In the
opposite limit, the four photons are in a four-particle entangled
state. Any intermediate case can be characterized by a single
parameter $\chi$, which is a function of $t_c^{ph}/\Delta t$. We
present a direct measurement of $\chi$ through a simple
experimental setup. The full theoretical analysis is also
provided.
\end{abstract}

\maketitle

\section{Introduction}

Spontaneous parametric down-conversion (SPDC) \cite{mandel,walls}
is a light amplification process that takes place in a non-linear
$\chi^{2}$ medium, where a photon from a pump field is converted
into two photons, usually called signal and idler, with energy and
momentum conservation. The signal and idler fields are therefore
strongly correlated in energy, emission times, polarization and
momentum. SPDC is a very convenient tool to produce entangled
states of photons, which have been used to test the foundation of
quantum physics and are at the heart of quantum information
processing and communications (see \cite{tw} for a review). In the
basic setup, the pump field is cw, and the output state is the
so-called two mode squeezed state (see e.g. \cite{walls} Eq.
(5.64)). When the pump intensity is low enough, the output state
is well described by a large vacuum component plus a two-photon
state, a {\em photon pair}. Recently, physicist have started to go
beyond this basic configuration. On the one hand, when the
classical pump field is no longer cw but pulsed. In this case, the
down-conversion process can take place only when a pump pulse is
"inside" the crystal, thus providing an information about the time
at which the down-converted photons are emitted. Of course, as a
counterpart, coherence is lost in the frequency domain, since a
pulsed pump field is not monochromatic. Non-trivial effects of a
pulsed pump have been predicted \cite{keller,grice} and observed
\cite{exper} for photon pairs. On the other hand, more efficient
sources and large pump intensities allow to produce an output
field where the four -and more- photons components are longer
negligible. The {\em four-photon component} of the field is of
interest in quantum optics \cite{wang} and in quantum information,
since four-qubit entanglement can be obtained
\cite{wein,howell02,Eibl 03}. But this component can be a nuisance
as well, for instance in quantum teleportation, because its
presence decreases the fidelity of the two-qubit Bell state
measurement \cite{HDR03} or in two-photon interference experiments
where it limits the visibility \cite{marcikic02}.

In this paper, we investigate both experimentally and
theoretically the physics of the four-photon component produced in
down-conversion, with a pulsed pump field. We start with a
qualitative description of what is to be expected. The two
meaningful quantities are the duration of the pump pulse $\Delta
t$ and the coherence time of the observed down-converted photons
$t_c^{ph}$. The characteristics of the four-photon state are
captured by the relation between $P_2$ and $P_4$, the
probabilities of creating 2, resp. 4 photons.

Let us consider first the limit $t_c^{ph}\ll\Delta t$. A large
number of independent SPDC processes can take place inside a pump
pulse \cite{comment1}. In this limit, any $2n$ photon state can be
satisfactorily described as n independent pairs. The probability
of creating $n$ pairs in a given pulse is described according to a
Poisson distribution of mean value $\mu$ $P_n=e^{-\mu}\mu^n/n!$,
as shown in Section \ref{secstate}. In particular, for small $\mu$
we have $P_4\approx \frac{P_2^2}{2}$, and the four-photon state
corresponds to two independent pairs, labelled $\ket{2\mbox{
EPR}}$.

The other limit, $t_c^{ph}\gg\Delta t$, can be achieved by the use
of femtosecond pump pulses and narrow filtering of signal and
idler photons. This condition is mandatory for all experiments
where photons created in different SPDC events must interfere at a
beam splitter, in order to preserve the temporal
indistinguishability \cite{zukowski95}. In this case, the emission
of a "second pair" is stimulated by the presence of the first one
\cite{Lamas01} leading to an entangled four-photon state
$\ket{4\mbox{ entg}}$ that cannot be described as two independent
pairs. Because of stimulated emission, we have $P_4 = {P_2}^2$.

In the present paper, we study the transition between these two
extreme situations. We define a parameter $\chi \in [0,1]$ that
allows to interpolate between the Poisson distribution and the
statistics arising from stimulated emission according to \ba
P_4&=&\frac{{P_2}^2}{2}(1+\chi)\,. \label{refchi}\ea In Section
\ref{secstate}, we present an intuitive description of the physics
in the language of quantum states. In Section \ref{secexp}, we
present a simple experiment that allows to measure $\chi$. In
Section \ref{secth}, we give the full quantum-optical formalism to
describe the four-photon component of the field, and work out an
approximate solution which agrees well with the experimental data.
In particular, we find that $\chi$ depends on $t_c^{ph}$ and
$\Delta t$ only through their ratio \ba r &=& t_c^{ph}/\Delta
t\,.\ea

\section{The four-photon state}
\label{secstate}

A coherent down-conversion process produces the two-mode squeezed
state \ba \ket{\Psi}&=&\sum_{n}\frac{T^n}{C}\,\ket{n;n}\ea where
$T\equiv \tanh \xi$ and $C\equiv \cosh \xi$ and $\xi$ is
proportional to the amplitude of the pump field (see \cite{walls}
Eq. (5.64)). The state $\ket{n;n}$ describes the field with $n$
photons in the signal mode and $n$ photons in the idler mode; in
particular, $\ket{2;2}\equiv\ket{4\mbox{ entg}}$ is the
four-photon state described in the introduction.

In the limit where we study photons whose coherence is much larger
than the time-bin (the width of the pump pulse), that is
$t_c^{ph}\gg\Delta t$, there is a unique coherent down-conversion
process taking place in the crystal for each pump pulse, and in
this case the four-photon component of the field is $\ket{4\mbox{
entg}}$. We expect $\chi= 1$.

If $t_c^{ph}<\Delta t$, a number $N\approx \Delta t/t_c^{ph}=1/r$
of independent SPDC processes can take place inside a pump pulse.
To simplify this discussion, we consider $N$ as an integer. The
state of the field reads then \ba \ket{\Psi}^{\otimes
N}&=&\frac{1}{C^N}\,\sum_n T^n\,\ket{2n}\,=\,\frac{1}{C^N}\,
\big(\ket{0}+T\ket{2}+T^2\ket{4}+...\big)\,, \label{psin}\ea where
the $\ket{2n}$ is the {\em un-normalized} superposition of all the
states containing $2n$ photons. Specifically:

The two-photon component $\ket{2}$ is the sum of the $N$ states
that describe "one pair in process $j$ and no pairs in the other
processes", that is
$\ket{2}=\ket{1;1}\otimes\ket{0}...\otimes\ket{0}\,+\,...\,+\,
\ket{0}\otimes\ket{0}...\otimes\ket{1;1}$. Since all the
components are mutually orthogonal, we have $\braket{2}{2}=N$. The
four-photon component $\ket{4}$ is the sum of two kind of terms:
(i) The $N$ components
$\ket{2;2}\otimes\ket{0}\cdots\otimes\ket{0}+\ket{0}\otimes\ket{0}\cdots\otimes\ket{2;2}$,
in which the second pair is created by stimulated emission; each
of these gives rise to the correlations of $\ket{4\mbox{ entg}}$.
(ii) The $N(N-1)/2$ components where one pair is produced in
process $j$ and another pair is produced in a different process
$j\,'$. Each of these gives rise to the correlations of
$\ket{2\mbox{ EPR}}$. Therefore $\braket{4}{4}=N(N+1)/2$, and by
normalizing this component we can say that the "four-photon state"
is \ba \ket{\Psi_4(N)}&=& \sqrt{\frac{2}{N+1}}\,\ket{4\mbox{
entg}}\,+ \,\sqrt{\frac{N-1}{N+1}}\,\ket{2\mbox{ EPR}}\,. \ea
Referring back to (\ref{psin}), we can compute the probabilities
of having two or four photon: $P_2(N)\,=\,
\frac{1}{C^{2N}}\,T^2\,N$, $P_4(N)\,=\, \frac{1}{C^{2N}}\,T^4\,
\frac{N(N+1)}{2}$. In the limit of very large number of
independent processes $N$, the usual argument leads to the Poisson
distribution \cite{notepoiss}. Now we have all the elements to
compute $\chi$ and relate it to the description of the four-photon
component. For simplicity, we put $C=1$. Then from (\ref{refchi})
we obtain $\chi=2P_4(N)/P_2(N)^2-1$ that is \ba \chi\,=\,
\frac{1}{N}\,=r&\mbox{
 for  }&t_c^{ph}<<\Delta t \ea and we can re-write the
four-photon state as \ba \ket{\Psi_4(\chi)}&=&
\sqrt{\frac{2\chi}{1+\chi}}\,\ket{4\mbox{ entg}}\,+
\,\sqrt{\frac{1-\chi}{1+\chi}}\,\ket{2\mbox{ EPR}}\,. \ea This
provides an intuition on the link between $\chi$, the experimental
parameter $r$ and the entanglement in the four-photon state.

\section{The experiment}
\label{secexp}

\begin{figure}[h]
\includegraphics[angle=0,width=12cm]{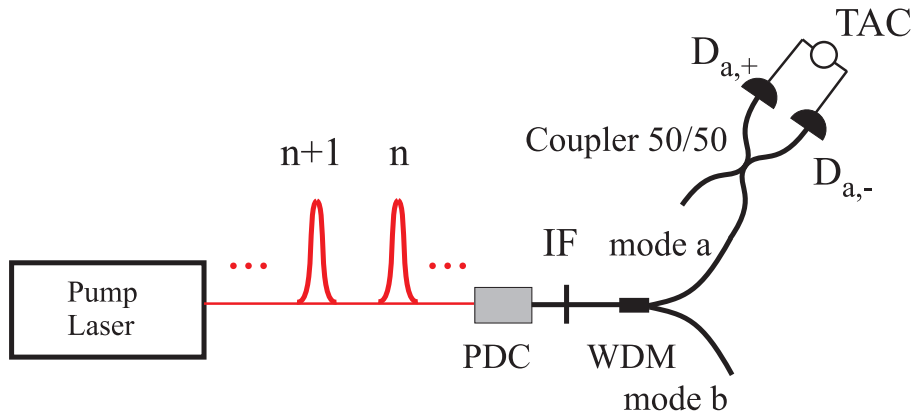}
\caption{Schematic of the experiment used to measure the
stimulation parameter $\chi$. See text for details.} \label{setup}
\end{figure}

A schematic of the experiment that measures $\chi$ is shown in
Fig. \ref{setup}.
A non-degenerate type I parametric down-converter is pumped by a
pulsed laser. PDC modes $a$ and $b$ are then separated
deterministically, via their different wavelengths. We ignore
photons in mode $b$; in mode $a$, we detect coincidence counts
between the outcomes of a passive coupler. This coincidence
measurement post-selects the events in which at least four photons
have been produced in the down-conversion processes. The idea is
to compare the events where four photons are created in the same
pump pulse with the events where one pair is created in one pulse
and another one in the next pulse. In the first case, we detect a
coincidence in a time window centered at $\Delta t_{det}=0$. The
coincidence count rate $R_0$ of this peak is proportional to
$\frac{1}{2}P_{4}$. The factor $\frac{1}{2}$ is the probability
that the two photons exit different modes of the beam splitter. In
the second case, we detect a coincidence with $\Delta
t_{det}=\Delta \tau _{laser}$ (time between 2 laser pulses). We
restrict ourself to the case when a photon created in pulse n is
detected by detector $D_{a,+}$ while photon created in pulse $n+1$
is detected by detector $D_{a,-}$. The coincidence count rate
$R_{side}$ is thus proportional to $\frac{1}{4}[P_2(I)]^2$.
Consequently, we have \ba \frac{R_0}{R_{side}} =
\frac{2P_{4}}{[P_2]^2}=\chi+1. \ea

Here is a description of the experimental setup. A mode-locked
femto-second laser, generating Fourier transform limited pulses at
710 nm, is used to pump a lithium triborate (LBO) type I non
linear crystal. The time between two subsequent laser pulses is
$\Delta \tau _{laser}=13ns$. Collinear non-degenerate photon pairs
at telecom wavelengths (1310 and 1550 nm) are created by
parametric down-conversion. The created photons are then coupled
into an optical fiber and separated deterministically with a
wavelength division multiplexer (WDM). We ignore the 1550 nm
photons and we send the 1310 nm photons to a 50-50 fiber coupler.
The two outputs of the coupler are connected to photon detectors,
labelled $D_{a,+}$ and $D_{a,-}$ in Fig. \ref{setup}. These
detectors are Ge and InGaAs avalanche photodiodes (APD),
respectively operating in Geiger mode. The Ge APD is used in
passive quenching mode, while the InGaAs is used in the so called
gated mode. In order to reduce the noise, the trigger for the
InGaAs APD is given by a coincidence between the Ge APD and a 1 ns
signal delivered simultaneously with each laser pulse. The signal
from one detector serves as START for a Time to Amplitude
Converter (TAC), while the signal from the other one serves as
STOP. We can thus measure the arrival times and see directly the
effect of stimulated emission when the photons are created in a
same pulse. An interference filter (IF) of different spectral
width $\Delta \lambda$ (5nm, 10nm, 40nm, FWHM) can be placed after
the crystal, in order to change the coherence length of the
down-converted photons. The coherence time (FWHM) is calculated
from $\Delta \lambda$ assuming a gaussian spectral transmission
for the IF: $t_{c}^{ph}=0.44\frac{\lambda^2}{c\Delta \lambda}$
\cite{comment0}. The gaussian transmission is a very good
approximation for $\Delta \lambda$ =5,10 nm, but it is less
accurate for $\Delta \lambda$=40 nm. The calculated $t_{c}^{ph}$
for $\Delta \lambda$=40 nm might therefore be underestimated. The
pump pulse duration $\Delta t$ can also be varied, and is measured
after the crystal with an auto-correlator. Note that the pump
pulses are no more completely Fourier transform limited after the
crystal, due to chromatic dispersion in the optical path. With the
different IF and the different pump pulse durations, it is thus
possible to vary the ratio $r=t_{c}^{ph}/\Delta t$.
\begin{figure}[h]
\includegraphics[angle=0,width=12cm]{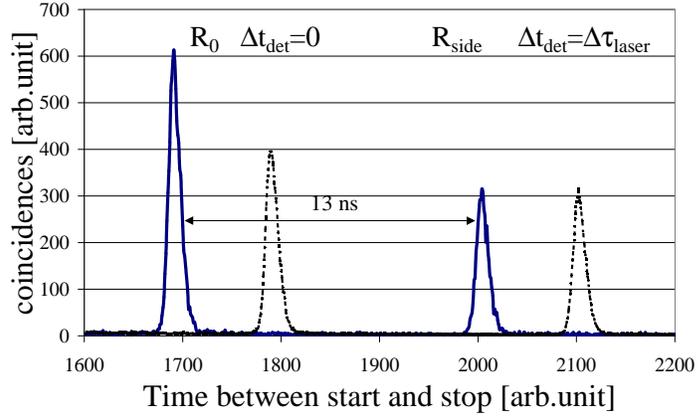}
\caption{TAC histogram for two different ratios r. The straight
line corresponds to $r$=2.5 (measured $\chi= 0.95$) and the dotted
line to $r$=0.2 (measured $\chi=0.3$) The two curves are
artificially shifted for clarity.} \label{TAC}
\end{figure}
Fig. \ref{TAC} shows a typical TAC histogram for two different
configurations (i.e. pump pulse duration and IF), leading to
different values of $r$. For each value of $r$, one can directly
measure the $\chi$ by comparing the number of coincidence counts
$R_{0}$ in the central peak and $R_{side}$ in the side peak.
\begin{figure}[h]
\includegraphics[angle=0,width=12cm]{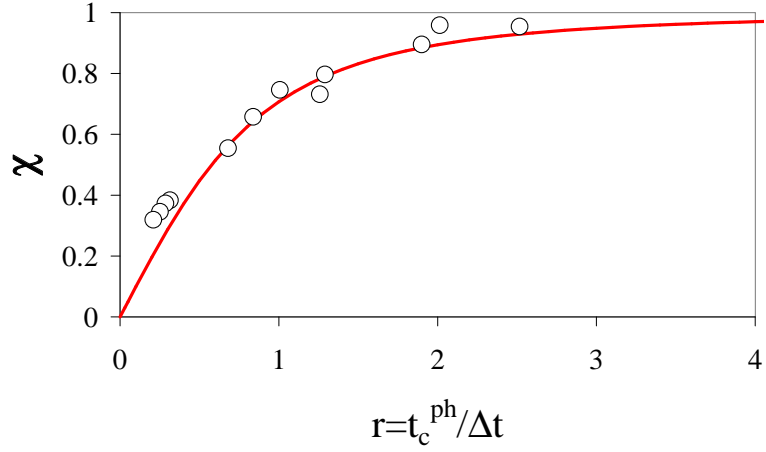}
\caption{Measured $\chi$ coefficient as a function of the ratio
$r$. The open circles are experimental points. The full line is a
plot of formula (\ref{chifexpl}).} \label{results}
\end{figure}
Fig. \ref{results} shows the measured $\chi$, as a function of the
ratio $r$. The theoretical prediction described later is \ba
\chi&=&\frac{r}{\sqrt{1+r^2}}\,, \label{chifexpl}\ea which by the
way reproduces the predictions of Section \ref{secstate} in the
limiting cases ($\chi=1$ for large $r$ and $\chi=r$ for small
$r$). The agreement of the data with this prediction is
satisfactory --- note in particular that there are no free
parameters in the model.

\section{Theoretical description}
\label{secth}

\subsection{State after down-conversion}
We want to describe the state of the e-m field produced as the
output of a down-conversion process driven by a classical field.
The scheme of the calculation is standard: the output state is \ba
\ket{\Psi}&=& e^{-i{\cal{H}}}\ket{0}\,\simeq \,
\big(\one-i{\cal{H}}-\demi{\cal{H}}^2\big)\ket{0}\,,
\label{psidef}\ea where ${\cal{H}}=\frac{1}{\hbar}\int H_{I}(t)dt$
and $H_{I}(t)$ is the Hamiltonian describing the down-conversion
process in the interaction representation. This scheme has been
applied in Refs. \cite{keller,grice} to the case in which the pump
field is not a continuous cw wave but a pulse of finite
spatio-temporal extension. In these Refs, the calculation was
limited to the two-photon term. Wang et al. \cite{wang} studied
the four-photon term, for degenerate type-I down-conversion. Here
we consider colinear emission (propagation along $z$) of
non-degenerate photons in a type-I crystal.

The Fourier-transform of the pump field is written
$\sqrt{I_p}\,\alpha(\omega)$. Following the same steps as in Refs.
\cite{keller,grice} we find ${\cal{H}}\,=\,
\sqrt{I}\,\left({\cal{A}}^{\dagger} \,+\,{\cal{A}}\right)$, where
$I$ is proportional to the intensity $I_p$ of the pump field, and
where ${\cal A}^{\dagger}$ is the two-photon creation operator \ba
{\cal A}^{\dagger}&=&\int d\omega_1\,d\omega_2
\,\alpha(\omega_1+\omega_2)\, \Phi(\omega_1,\omega_2)\,
a^{\dagger}(\omega_1) a^{\dagger}(\omega_2)\,.\label{cala}\ea Here
appears the phase-matching function $\Phi(\omega_1,\omega_2)$,
peaked around $\omega_1\approx \Omega_1$ and $\omega_2\approx
\Omega_2$ with $\Omega_1 \neq \Omega_2$ \cite{phasematch}. In the
following for conciseness we shall write \ba {\cal
P}(\omega_1,\omega_2)&=& \alpha(\omega_1+\omega_2)\,
\Phi(\omega_1,\omega_2)\,. \label{calp}\ea Now we must insert
${\cal{H}}$ into (\ref{psidef}). Since ${\cal A}\ket{0}=0$ and
${\cal A}{\cal{A}}^{\dagger}\ket{0}\,\propto\,\ket{0}$, after
removing the vacuum component that obviously does not contribute
to the detection, one finds \ba \ket{\Psi}&\simeq &
i\sqrt{I}\,{\cal A}^{\dagger}\ket{0}+\frac{I}{2}\, {{\cal
A}^{\dagger}}^2\ket{0}\,. \label{psiprep}\ea

\subsection{Probabilities and the observed $\chi$}

From $\ket{\Psi}$ produced at the down-conversion, we can
calculate the probabilities $P_2=I\sandwich{0}{{\cal A}{\cal
A}^{\dagger}}{0}$ and $P_4=\frac{I^2}{4} \sandwich{0}{{\cal
A}^2{{\cal A}^{\dagger}}^2}{0}$ of producing exactly two,
respectively exactly four, photons. For this, one makes use of \ba
\left[a(\omega),a^{\dagger}(\omega')\right]
&=&\one\,\delta(\omega- \omega')\, \label{commut}\ea and of a
corollary of this commutation rule, that reads \ba
\sandwich{0}{a(\omega'') a(\omega''')a^{\dagger}(\omega)
a^{\dagger}(\omega')}{0}\nonumber \\ =
\delta(\omega-\omega'')\delta(\omega'-\omega''')\nonumber \\+
\delta(\omega-\omega''')\delta(\omega'-\omega'')\,.
\label{coro}\ea For $P_2$, the result is formally $P_2=I\int
d\omega_1 d\omega_2 \left[|{\cal P}(\omega_1,\omega_2)|^2+{\cal
P}(\omega_1,\omega_2) {\cal P}^{*}(\omega_2,\omega_1)\right]$.
However, the second term of the sum is always zero because of the
phase-matching condition (non-degenerate photons): in fact, the
ranges of frequencies in which the first and the second argument
of $\Phi$, and consequently of ${\cal P}$, lead to a non-zero
contribution do not overlap. From now onwards, we simply consider
that $\omega_1$ and $\omega_2$ are different, that is
$\left[a(\omega_1),a^{\dagger}(\omega_2)\right]=0$. Here is the
point when our calculation differs from the one of Ref.
\cite{wang}. In conclusion, we have obtained \ba P_2&=& I\,\int
d\omega_1 d\omega_2 |{\cal
P}(\omega_1,\omega_2)|^2\,\equiv\,I\,J_2\,. \label{prob2}\ea To
calculate $P_4$, formula (\ref{coro}) is applied to the operators
acting on both modes $a$ and $b$, and one obtains \ba
P_4&=&\frac{1}{2}\,I^2\, \left[(J_2)^2\,+\,J_{4}\right]
\label{prob4}\ea where, writing $d\underline{\omega}=d\omega_1
d\omega_bd\omega_1' d\omega_2'$,  \ba J_4&\equiv&\int
d\underline{\omega} {\cal P}(\omega_1,\omega_2) {\cal
P}(\omega_1',\omega_2') {\cal P}^*(\omega_1',\omega_2) {\cal
P}^*(\omega_1,\omega_2') \nonumber\\ &=&\int d\omega_2
d\omega_2'\,\left|\int d\omega_1 {\cal P}^*(\omega_1,\omega_2')
{\cal P}(\omega_1,\omega_2) \right|^2\label{i4}\ea One can verify
that $0\leq J_4\leq (J_2)^2$ \cite{wang}. By comparison with eq.
(\ref{refchi}) we find $\chi_0 \,\equiv\, J_4/(J_2)^2$; here, the
suffix "0" means that this is the value of $\chi$ in the absence
of any filtering after the SPDC.

Now we must move and describe our experiment. Two approaches are
possible. The "brute force" approach consists in effectively
describing all the details of the experiment: write the pump pulse
consisting of two well-separated pulses (this configuration gives
similar experimental predictions as the one presented in section
II, but is easier to compute), have mode $a$ evolve through the
coupler, and finally compute the coincidence rate at the
detectors. This calculation is lengthy, although not devoid of
interest for the theorist; we give it as an Appendix. The second
approach is more clever: we know that the experiment measures
$\chi$ for a single pump pulse $g(\omega)$, in the presence of an
interference filter. We can then apply all that we have just done
to find immediately \ba \chi &=& J_4^F/ \big(J_2^F\big)^2
\label{chif}\ea where, writing $G(x,y)=g(x+y)\Phi(x,y)$ and
$F(\omega)$ for the intensity profile of the filter, we have \ba
J_2^F&=&\int d\omega_1
d\omega_2 F(\omega_1) |G(\omega_1,\omega_2)|^2\,, \label{j2f}\\
J_4^F&  = &\int d\underline{\omega}\,F(\omega_1)F(\omega_1')\,
G(\omega_1,\omega_2) G(\omega_1',\omega_2')
G^*(\omega_1',\omega_2) G^*(\omega_1,\omega_2')\,. \label{j4f}\ea
Obviously, in the Appendix we obtain the same result.

\subsection{Explicit estimate}

We have just found the general formulae that describe the quantity
$\chi$ measured in our experiment. In this paragraph, we solve
explicitly (\ref{j2f}) and (\ref{j4f}) using some crude
approximations; the final result will be formula (\ref{chifexpl})
for $\chi$, that has been shown to fit the experimental data.

We make the following hypotheses:\\
(i) The pump pulse is Fourier-transform limited:
$g(\omega)=\left(2\pi\Delta_p\right)^{1/4}\,e^{-(\omega-\Omega_p)^2/4\Delta_p^2}$.\\
(ii) The filter has a gaussian profile:
$F(\omega)=\left(2\pi\Delta_F\right)^{1/2}\,e^{-(\omega-\Omega_1)^2/2\Delta_F^2}$.\\
(iii) The coherence time of the photons is uniquely determined by
the width of the filter: $t_c^{ph}=\Delta_F^{-1}$. Therefore we
replace $\Phi(\omega_1,\omega_2)$ by 1, that is,
$G(\omega_1,\omega_2)$ by $g(\omega_1+\omega_2)$.

The big advantage of this set of hypotheses however is that we are
left with two meaningful quantities: $\Delta_p$ and $\Delta_F$,
whose inverses are the coherence times of the pump and of the
photons in mode $a$. Plausible physical arguments will allow us to
get rid of all these hypotheses at the end of the calculation.

The calculation of $J_2^F$ is straightforward: via the change of
variables $(\omega_1,\omega_2)\rightarrow
(\omega_1,\xi=\omega_1+\omega_2)$, the double integral factorizes
into the product of two normalized gaussians, so $J_2^F=1$. This
implies $\chi=J_4^F$. The calculation of this integral is longer
but not difficult: by the usual technique of square completion,
one first arranges the terms in order to integrate out $\omega_2$
and $\omega_2'$; the result of this procedure, writing
$\omega_1=x$ and $\omega_1'=y$, is \ba
J_4^F\,=\,\frac{1}{2\pi\Delta_F}\int
dxdy\,\exp\left[-\demi\left(\frac{x^2+y^2}{\Gamma^2}
-\frac{xy}{\Delta_p^2}\right)\right]\ea with $\frac{1}{\Gamma^2}=
\frac{1}{\Delta_{F}^2} + \frac{1}{2\Delta_{p}^2}$. Using square
completion again, one can integrate on $y$ first and on $x$ later.
The result is
$J_4^{F}=\frac{\Gamma}{\Delta}\,\frac{\Gamma'}{\Delta}$ with
$\frac{1}{\Gamma'^2}\,=\, \frac{1}{\Gamma^2} -
\frac{\Gamma^2}{4\Delta_{p}^2}$. A little more algebra leads to:
\ba \chi\,=\,\frac{\tilde{r}}{\sqrt{1+\tilde{r}^2}}\,, \mbox {
with }\, \tilde{r}\,=\,\frac{\Delta_p}{\Delta_F}\,.
\label{chifexpl1}\ea From this simple result, we guess the general
result (\ref{chifexpl}) by identifying $\tilde{r}$ with $r$. This
step is motivated by the following considerations. One the one
hand, since we took a Fourier-transformed limited pulse for the
calculation, the coherence time of the pump $1/\Delta p$ is equal
to the pulse duration, which we know to be the relevant quantity
\cite{comment1}. On the other hand, $1/\Delta F$ is the coherence
time of the down-converted photons, as long as the filter
bandwidth is much smaller than the bandwidth of the unfiltered
photons. When this condition is no longer satisfied, the relevant
physical parameter is of course $t_c^{ph}$.

\section{Conclusion}
We studied the physics of four-photon states in pulsed parametric
down-conversion. The parameters of the experiment determine to
which extent the four-photon state exhibits four-photon
entanglement, or can be rather described as two independent pairs.
Any intermediate situation is quantified by a single parameter
$\chi$ that depends only on the ratio between the coherence time
of the created photons and the duration of the pump pulse. A
simple experiment to measure $\chi$ has been realized. A
theoretical model based on the standard formalism of quantum
optics has been derived, that fits well the experimental data.
Beside its fundamental aspect, this work is of practical interest
in quantum optics, because it provides a simple mean to quantify
the "coherence" of four photons states which is important for
experiments such as quantum teleportation where independent
photons must overlap at a beam splitter.\\ \emph{Note :} For
related independent works on the statistics of photon numbers in
down-conversion, see \cite{stat}\\

The authors would like to thank Claudio Barreiro and Jean-Daniel
Gautier for technical support. Financial support by the Swiss NCCR
Quantum Photonics, and by the European project RamboQ is
acknowledged.

\section{Appendix}

In this appendix, we want to re-derive formulae (\ref{j2f}) and
(\ref{j4f}) with a full calculation of the experiment sketched in
Fig. \ref{setup}.

For the experiment we are going to consider, the classical pump
field will be composed by two identical pulses separated by a time
$\tau$. This configuration leads to the same experimental results
as the one presented before and is easier to compute. If
$g(\omega)$ is the Fourier transform (FT) of each pulse, the FT of
the pump field is then simply
$\alpha(\omega)=g(\omega)(1+e^{i\omega\tau})$. To avoid
multiplying notations, we keep ${\cal P}$ as in (\ref{calp}) for
this explicit form of the pump field: \ba {\cal
P}(\omega,\tilde{\omega})&=&(1+e^{i(\omega+\tilde{\omega})\tau})\,
g(\omega+\tilde{\omega})\, \Phi(\omega,\tilde{\omega})\,\nonumber
\\&=&\,
(1+e^{i(\omega+\tilde{\omega})\tau})\,G(\omega,\tilde{\omega})\,.
\label{fexpl} \ea

\subsection{Evolution}

As discussed, the photons are separated according to their
frequency. Those whose frequency is close to $\Omega_1$ (resp.
$\Omega_2$) are coupled into the spatial mode $a$ (resp. $b$).
Photons in mode $b$ do not undergo any evolution, while mode $a$
evolves through a 50-50 coupler into modes $c=(a,+)$ and $d=(a,-)$
according to\ba a^{\dagger}(\omega)&\longrightarrow&
\frac{c^{\dagger}(\omega)+ i\,d^{\dagger}(\omega)}{\sqrt{2}}\,.
\label{coupler}\ea Inserting this evolution into (\ref{psiprep}),
the state at the detection reads \ba \ket{\Psi}_{det} &\simeq &
i\,{\cal K}\ket{0}+\frac{1}{2}\, {{\cal K}}^2\ket{0}
\label{psidet}\ea where, omitting the multiplicative constant
$\sqrt{I/2}$, we have written \ba {\cal K}&=&\int
d\omega\,d\tilde{\omega} \,{\cal P}(\omega,\tilde{\omega})\,
\left[c^{\dagger}(\omega)+i d^{\dagger}(\omega)\right]\,
b^{\dagger}(\tilde{\omega})\,.\label{cala}\ea

\subsection{Detection (I): generalities}

We turn now to the detection. The experiment that we are
describing involves the detection of two-photon coincidences. Let
$D_c$ and $D_d$ be the two detectors that monitor modes $c$ and
$d$. The probability of detecting a coincidence between the two
events "photon detected in $D_j$ at the time $t_j$", $j=c,d$, is
given by \ba P_{D_c,D_d}(t_c,t_d)&=&||
E^{(+)}_{c}(t_c)\,E^{(+)}_{d}(t_d)\,\ket{\Psi}_{det} ||^2
\label{coinc}\ea where \ba E^{(+)}_{c}(t_c)\,&=&\,\int
d\nu\,f_c(\nu)\,c(\nu)\,e^{-i\nu t_c}\,\nonumber\\
E^{(+)}_{d}(t_d)\,&=&\,\int d\nu\,f_d(\nu)\,d(\nu)\,e^{-i\nu t_d}
\ea are the positive frequency component of the electric field at
time $t_j$ in the mode detected by detector $D_j$ \cite{note0},
weighted by the amplitude of the filter $f_j(\nu)$ put in front of
each detector. From now on, as in our experiment where the filter
was actually put before the WDM, we consider $f_c=f_d=f$; the
intensity shape of the filter is $F(\omega)=|f(\omega)|^2$.

Now we insert (\ref{psidet}) into (\ref{coinc}). As expected, the
term linear in ${\cal K}$ gives no contribution: if there is just
one photon in modes $c$ or $d$, no coincidence count can be
obtained. Similarly, no contribution comes from the terms of the
form $c^{\dagger}(\omega)c^{\dagger}(\omega')$ and
$d^{\dagger}(\omega)d^{\dagger}(\omega')$ in ${\cal K}^2$, where
both photons are found in the same mode after the coupler. In the
calculation of the non-zero terms, we systematically omit
multiplicative constants from now on. Using (\ref{commut}) for
modes $c$ and $d$, one finds:
\ba E^{(+)}_{c}(t_c)\, E^{(+)}_{d}(t_d)\,\ket{\Psi}_{det}= \int
d\underline{\omega}\, e^{-i(\omega t_c+\omega' t_d)}\,\,f(\omega)
\,f(\omega') {\cal P}(\omega,\tilde{\omega}) {\cal
P}(\omega\,',\tilde{\omega}\,') \,b^{\dagger}(\tilde{\omega})
b^{\dagger}(\tilde{\omega}\,') \ket{0}\,, \ea
where we have written $d\underline{\omega}=
d\omega\,d\omega\,'\,d\tilde{\omega}\,d\tilde{\omega}\,'$. The
probability $P_{D_c,D_d}(t_c,t_d)$ is the square modulus of this
expression. Using (\ref{coro}) for mode $b$, one finds
\ba P_{D_c,D_d}(t_c,t_d)= \int d\underline{\omega}\,
e^{-i[(\omega-\hat{\omega}) t_c+(\omega'-\hat{\omega}')
t_d)]}\,\big[K(\omega,\omega',\hat{\omega},
\hat{\omega}';\tilde{\omega},\tilde{\omega}\,') +
K(\omega,\omega',\hat{\omega}',
\hat{\omega};\tilde{\omega},\tilde{\omega}\,')\big]
\label{probatt}\ea
where we have written $d\underline{\omega}=
d\omega\,d\omega\,'\,d\hat{\omega}\,d\hat{\omega}\,'\,d\tilde{\omega}\,d\tilde{\omega}\,'$
and  
\ba K(\omega,\omega',\hat{\omega},
\hat{\omega}';\tilde{\omega},\tilde{\omega}\,')&=&f(\omega)
\,f(\omega')\,f(\hat{\omega}) \,f(\hat{\omega}')\, {\cal
P}(\omega,\tilde{\omega}) \,{\cal P}(\omega\,',\tilde{\omega}\,')
{\cal P}^*(\hat{\omega},\tilde{\omega})\, {\cal
P}^*(\hat{\omega}',\tilde{\omega}\,')\,.\ea
 The detection rate
(counts per pulse) is obtained by integrating
$P_{D_c,D_d}(t_c,t_d)$ over the time-resolution of the detector
$2\Delta T$: \ba R(T_c,T_d)&=&\int_{T_{c,d}\pm\Delta T}\,dt_c\,
dt_d\,P_{D_c,D_d}(t_c,t_d)\,. \label{rate}\ea The time-resolution
must be longer than the coherence time of the photons (otherwise,
the selected modes cannot be seen), and shorter than the spacing
between the pulses, to allow the resolution in time-bins. As the
result of this integration, $R(T_c,T_d)$ has the same form as
$P_{D_c,D_d}(t_c,t_d)$ given in (\ref{probatt}), via the
replacements \ba e^{-i(\omega-\hat{\omega})t_c}&\rightarrow&
\Delta T\,e^{-i(\omega-\hat{\omega})T_c}\,\mbox{sinc}[(\omega-\hat{\omega})\Delta T]\\
e^{-i(\omega'-\hat{\omega}')t_d}&\rightarrow& \Delta
T\,e^{-i(\omega'-\hat{\omega}')T_d}\,\mbox{sinc}[(\omega'-\hat{\omega}')\Delta
T]\,.\ea Even if it seems redundant here, for subsequent ease, it
is convenient to write down explicitly \ba R(T_c,T_d)&=&
R_1(T_c,T_d)+R_2(T_c,T_d)\ea where we have defined, writing
$S(x)=\Delta T\mbox{sinc}(x\Delta T)$:
\begin{widetext}
\ba R_1(T_c,T_d)&=&  \int d\underline{\omega}\,
e^{-i[(\omega-\hat{\omega}) T_c+(\omega'-\hat{\omega}')
T_d)]}\,K(\omega,\omega',\hat{\omega},
\hat{\omega}';\tilde{\omega},\tilde{\omega}\,')\,
S(\omega-\hat{\omega})S(\omega'-\hat{\omega}')\,,\\
R_2(T_c,T_d)&=& \int d\underline{\omega}\,
e^{-i[(\omega-\hat{\omega}) T_c+(\omega'-\hat{\omega}')
T_d)]}\,K(\omega,\omega',\hat{\omega}',
\hat{\omega};\tilde{\omega},\tilde{\omega}\,')\,
S(\omega-\hat{\omega})S(\omega'-\hat{\omega}')\,. \ea
\end{widetext}

\subsection{Detection (II): meaningful times}

We have just found an expression for $R(T_c,T_d)$. Now, recall
that the first time-bin is defined by $T=0$, the second time-bin
is defined by $T=\tau$. Therefore, we expect $R(T_c,T_d)$ to be
significantly different than 0 only when $T_c$ and $T_d$ take the
values $0$ or $\tau$. In particular, the counting rate in the
central peak is $R_c=R(0,0)+R(\tau,\tau)$, while
$R_{lat}=R(0,\tau)=R(\tau,0)$ are the counting rates in each of
the lateral peaks. We want to recover all these results out of our
general formula. In addition, we shall have manageable expressions
for $R_c$ and $R_{lat}$, allowing a fit of the experimental data.

Let us start with a qualitative argument, that is enough for our
purposes. Recall the expression (\ref{fexpl}) of ${\cal P}$. The
spectral width of $g$ is larger than (equal to, for
Fourier-transform limited pulses) $\frac{1}{\Delta t}$, which in
turn is much larger than $\tau^{-1}$ since we want two
well-separated pulses. This means that $g(\omega)$ is almost
constant in a frequency range of width $\tau^{-1}$. The
phase-matching function is also constant over such ranges, because
its typical width is the inverse of the coherence time $t_c^{ph}$
of the down-converted photons. Now, suppose that $T_c$ and $T_d$
are $0$ or $\tau$. If one inserts (\ref{fexpl}) into the
expression for $R(T_c,T_d)$ and develops the products, one finds
that $R(T_c,T_d)$ is a sum of terms that are the product of $g$,
$\Phi$ and a phase factor of the form $e^{i\Omega\tau}$, with
$\Omega$ some algebric sum of the $\omega$'s. The arguments above
prove that this phase fluctuates very rapidly in the frequency
space, unless $\Omega\simeq 0$. Therefore, when we integrate over
the $\omega$'s, all the terms will average to 0 but those whose
phase factor is 1. Moreover, by direct check one can easily get
convinced that if either of $T_c$ or $T_d$ is equal to a time when
no photon is expected, say $\frac{\tau}{2}$ or $2\tau$, then no
phases can be erased: the coincidence rate becomes zero. In
summary, the first step to simplify $R(T_c,T_d)$ consists in
writing down explicitly all the terms, and keep only those terms
whose phase factor is 1. From now onwards, we admit that $T_c$ and
$T_d$ are either $0$ or $\tau$.

Having erased terms that fluctuate as $\tau^{-1}$ in the frequency
space, a further simplification is possible. The argument of the
cardinal sine functions is $(\omega-\hat{\omega})\Delta T\sim
\Delta T/t_c^{ph}$. But as we said, $\Delta T$ must be larger than
$t_c^{ph}$, otherwise the photon cannot be seen by the detector.
Therefore the cardinal sine will only be significant if
$\omega\approx\hat{\omega}$, that is, we can replace
$\mbox{sinc}[(\omega-\hat{\omega})\Delta T]$ with $\frac{1}{\Delta
T}\,\delta (\omega-\hat{\omega})$.

All this simplification procedure is just a matter of patience.
One finds that $R_1(0,0)=R_1(0,\tau)=R_1(\tau,0)=
R_1(\tau,\tau)\equiv (J_2^F)^2$ given in (\ref{j2f}), while
$R_2(0,0)= R_2(\tau,\tau)\equiv J_4^F$ given in (\ref{j4f}), and
$R_2(0,\tau)=R_2(\tau,0)=0$. In conclusion, the detection rates in
the central peak and in each of the lateral peaks are: \ba
R_c&=&R(0,0)+R(\tau,\tau)\,=\,2[(J_2^F)^2+J_4]\,,\\
R_{lat}&=&R(0,\tau)\,=\,R(\tau,0)\,=\,(J_2^F)^2\,, \ea and the
ratio $R_c/2R_{lat}$ is equal to $1+\chi$ as announced.


\begin{references}
\bibitem{comment1} In this context, it becomes clear why we said that the duration of the pump
pulse and not its coherence time is important parameter. One could
indeed imagine a pulse with $\Delta t \gg t_c^{ph}$ whose
coherence time is much shorter than $t_c^{ph}$ (think for instance
to a femto-second pulse that experienced chromatic dispersion). In
this case, photons can be created in the whole duration of the
pulse and consequently there are many temporal modes of the
created photons inside the pump pulse, even if its coherence time
is shorter than $t_c^{ph}$
\bibitem{comment0}
To find the factor $\frac{2ln(2)}{\pi}=0,44$, we start by
$\sigma_t \sigma_\nu=\frac{1}{4\pi}$ where $\sigma_t$ and
$\sigma_\nu$ are the standard deviation of the temporal and
spectral intensities. The FWHM are then calculated with:
$t_c^{ph}=2\sqrt{2\ln{2}}\sigma_t$.
\bibitem{phasematch} We don't need the explicit formula for this function
in what follows; just note that, if we can neglect the
$\omega$-dependence of the term $A_m$ (that is, chromatic
dispersion) and the $z$-dependence of the non-linear
susceptibility $\chi^{(2)}$, the phase-matching function becomes
equal to
$\Phi(\omega_1,\omega_2)\,=\,\mbox{sinc}\Big[\big(K_1(\omega_1-\Omega_1)+
K_2(\omega_2-\Omega_2)\big)L\Big]$, see eq. (11) in \cite{grice}
for the exact definition of the functions $K_1$ and $K_2$.
\bibitem{note0} Here we write $t_j$ instead of
$t_j-\frac{z_j}{c}$, $z_j$ being the distance from the crystal to
the detector. This amounts to a choice of the origin of time,
which is irrelevant because experimentally we can synchronize at
will the time-windows of the detectors. Moreover, we neglect the
$\nu$-dependence of the field amplitude, that we set to 1 because
we don't measure absolute intensities.
\end{references}
\end{document}